\documentclass[]{jfm}
\usepackage{natbib}
\usepackage{amsmath}
\usepackage[dvips]{graphicx}
\usepackage{psfrag}

\begin{document}

\title[Flow Control: Money versus Time]{Money versus Time: \\
Evaluation of Flow Control in Terms of Energy Consumption and Convenience}

\author[B. Frohnapfel, Y. Hasegawa \& M. Quadrio]{
By B\ls E\ls T\ls T\ls I\ls N\ls A\ls\ns F\ls R\ls O\ls H\ls N\ls A\ls P\ls F\ls E\ls L$^1$, \ls
Y\ls O\ls S\ls U\ls K\ls E\ls\ns H\ls A\ls S\ls E\ls G\ls A\ls W\ls A$^{1,2}$\ls \\
\and
M\ls A\ls U\ls R\ls I\ls Z\ls I\ls O\ls \ns Q\ls U\ls A\ls D\ls R\ls I\ls O$^3$
}
\affiliation{
$^1$ Center of Smart Interfaces, TU Darmstadt \\ Petersenstr. 32, 64287 Darmstadt, Germany \\
$^2$ Department of Mechanical Engineering, The University of Tokyo \\ Hongo 7-3-1, Bunkyo-ku, Tokyo 113-8656, Japan\\
$^3$Dipartimento di Ingegneria Aerospaziale del Politecnico di Milano \\ via La Masa 34, 20156 Milano, Italy
}

\maketitle

\begin{abstract}
Flow control with the goal of reducing the skin friction drag on the fluid-solid interface is an active fundamental research area, motivated by 
its potential for significant energy savings and reduced emissions in the transport sector. Customarily, the performance of drag reduction techniques in internal flows is evaluated under two alternative flow conditions, i.e. at constant mass flow rate or constant pressure gradient. Successful control leads to reduction of drag and pumping power within the former approach, whereas the latter leads to an increase of the mass flow rate and pumping power. In practical applications, however, money and time define the flow control challenge: a compromise between the energy expenditure (money) and the corresponding convenience (flow rate) achieved with that amount of energy has to be reached so as to accomplish a goal which in general depends on the specific application.
Based on this idea, we derive two dimensionless parameters which quantify the total energy consumption and the required time (convenience) for transporting a given volume of fluid through a given duct. Performances of existing drag reduction strategies as well as the influence of wall roughness are re-evaluated within the present framework; how to achieve the (application-dependent) optimum balance between energy consumption and convenience is addressed. It is also shown that these considerations can be extended to external flows.
\end{abstract}

\section{Introduction}
\label{intro}

Global issues such as depletion of energy resources and deterioration of natural environment face the modern society with the task of saving energy without compromising quality of life. Flow control opens up new possibilities to design and improve thermal-fluid systems. In particular, turbulent skin-friction drag reduction, which is the focus of the present paper, could have tremendous impacts on economy and ecology through innovation of energy-efficient fluid transport systems such as pipelines, high-speed vehicles, and so forth, where consideration of energy and of its costs are becoming increasingly important in flow control. 

Up to now, various drag-reducing techniques, applied to the canonical turbulent channel or pipe flows, have been explored either through Direct Numerical Simulation (DNS) of the Navier--Stokes equations, or by laboratory experiments. In DNS, their control performance has been evaluated while keeping constant in time either the flow rate (CFR) or -- less often -- the pressure gradient (CPG). Several examples of the former approach exist, starting from \cite{jung-mangiavacchi-akhavan-1992} and \cite{choi-moin-kim-1994}, whereas the CPG approach is for example exemplified in \cite{quadrio-ricco-2011}. The laboratory counterpart of such simulations is a set-up in which the flow rate is constantly measured and adjusted to remain at a fixed value, or -- alternatively -- in which a constant pressure overhead drives the flow. A pump with a characteristic operating curve would correspond to neither of these states, with flow rate and prevalence both changing as the operating point is changed.

Imposing a constraint on either the flow rate or the pressure gradient is advantageous from the standpoint of the mathematical formulation of the flow control problem. Taking advantage of the CFR condition, for example, \cite{fukagata-iwamoto-kasagi-2002} have derived a mathematical relationship between the wall friction and various dynamical contributions, later extended by \cite{marusic-joseph-mahesh-2007} to the CPG condition.
More recently, \cite{bewley-2009} and \cite{fukagata-sugiyama-kasagi-2009} proved for straight ducts that under CFR the total energy consumption (sum of pumping and control energy) is minimized when the flow becomes laminar, indicating that the ultimate goal in drag reduction control for energy saving is the complete relaminarization of the flow. For arbitrary-shaped ducts, where secondary flow may be generated, \cite{fukagata-sugiyama-kasagi-2009} demonstrated that Stokes flow minimizes total power consumption.

Under the CFR condition, a successful drag-reducing technique effectively reduces friction drag, which immediately translates into a reduction of the pumping energy. Since by definition active control  requires additional energy, recent investigations are shifting their emphasis from the raw drag reduction rate $R$ to the net energy saving rate $S$, which corresponds to the reduction rate of the total energy consumption, and also the gain $G$, defined as the ratio between the reduction of pumping energy and the control energy input. Typical values of $R$, $S$ and $G$ achieved by existing active controls are summarized by \cite{kasagi-hasegawa-fukagata-2009}. Similarly, \cite{hoepffner-fukagata-2009} propose to evaluate flow control performance under CFR by using a power plane, in which the ordinate represents the total energy consumption, whereas the abscissa is the difference between the pumping and control energy. One important drawback of imposing the CFR constraint, however, is that the wall shear stress, which is a dominant factor in near-wall turbulence dynamics, is changed due to the applied control, so that it is difficult to extract the essential effects of a control input itself due to superimposed Reynolds number effects. When the CPG condition is used, on the other hand, friction drag is indeed unchanged by design, and 'drag reduction' manifests itself through an increase of the flow rate, which implies an increase in the power required to drive the flow.

From the industrial standpoint, the interest for drag-reduction strategies is accompanied by difficulties of transferring techniques developed in academic research to practical applications. \cite{spalart-mclean-2011} have recently highlighted the need for the flow control community to be aware of energy issues when proposing active techniques, and advocated the importance of taking into account the combined effects of different factors on the overall efficiency (or cost) of a system. In the present problem, two factors play pivotal roles: flow rate and pressure gradient. The flow rate, kept constant in the CFR approach, is an essential consequence  of using a duct to transport a given amount of fluid through the duct over a certain distance, and we term it {\em convenience}. The pressure gradient, kept constant in the CPG approach, bears a direct relationship to the {\em energy consumption} required to achieve that convenience. In real-life applications using flow control, minimizing energy consumption for a given flow rate (the CFR approach) and maximizing convenience for a fixed energy consumption (the CPG approach) are only two of the many possible strategies conceivable to balance the two intervening factors. In general, the designer of a fluidic system would consider both convenience and energy requirements, the relative value and cost of which will depend on the specific application. The optimal use of a control technique is the one that achieves maximum value at minimum cost, as determined by the designer.

The aim of the present paper is thus to develop a conceptual framework where an unequivocal assessment of (non necessarily active) flow control techniques against whatever application-dependent value-for-money considerations will be possible. A new evaluation plane is proposed in which both quantities, i.e. energy consumption and convenience, are simultaneously and explicitly considered. This new plane can be viewed as an improved version of the familiar $C_f - Re$ plane, which describes in a dimensionless way how the flow rate and the pressure gradient required to achieve that flow rate are related. In the new plane, an analogous non-dimensional description relates the flow rate and the energy expenditure required to achieve that flow rate, possibly including control energy. Re-evaluating existing drag-reduction data by taking advantage of this plane will give us new insight on their performance and on the way we have to go for flow control techniques to become reality in applications.

\section{Internal flows}

\subsection{Dimensional analysis}

\begin{figure}
\centering
\includegraphics[width=0.55\textwidth]{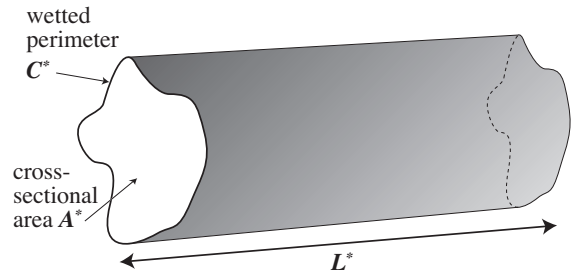}
\caption{Duct with constant cross-sectional area in the streamwise direction $x$.}
\label{fig:duct}
\end{figure}

We shall consider a given fluid volume $V^*_f$ which has to be transported through a duct of streamwise length $L^*$, as shown in Fig.~\ref{fig:duct}, by means of a pressure gradient. (The asterisk represents dimensional quantities throughout this paper.) The flow is assumed to be fully developed. The cross sectional area $A^*$ and the wetted perimeter $C^*$ of the duct do not vary along the streamwise direction $x$. The total wetted area is given by $C^*L^*$. The hydraulic diameter $D^*$ is defined as $D^* = 4A^* / C^*$.

According to the streamwise force balance, the time-averaged pressure gradient $- \Pi^*$ and the wall-shear-stress $\tau^*_w$ are related as:
\begin{equation}
\label{eq:tau_dp}
- \Pi^* A^* = \tau_w^* C^* .
\end{equation}

The time-averaged pumping power $P^*_p$ per unit wetted area is given by:
\begin{equation}
P^*_p = - \Pi \frac{A^*}{C^*} U^*_b = \tau^*_w U^*_b.
\label{eq:P_p}
\end{equation}
Here, $U^*_b$ is the bulk mean velocity, i.e. the time- and volume-averaged streamwise velocity. Strictly speaking, Eq.~(\ref{eq:P_p}) holds only if either the spatial average of the pressure gradient or the volume average of the streamwise velocity is not time-dependent. For other cases additional terms arise, which are however negligibly small unless the pressure gradient is actively varied, so that the correlation between the spatial average of the pressure gradient and the bulk mean velocity becomes significant.

The operating time $T^*$ required to transport the fluid volume $V^*_f$ across the duct is given by
$ T^* = V^*_f / (A^* U^*_b)$,
so that the pumping energy $E^*_p$ per unit wetted area can be written as:
\begin{equation}
\label{eq:E_p}
E^*_p = P^*_p T^* = \tau^*_w \frac{V^*_f}{A^*}.
\end{equation}
The dimensionless friction coefficient $C_f$ is defined as:
\begin{equation}
\label{eq:C_f}
C_f  = \frac{\tau^*_w}{\frac{1}{2}\rho^* {U^*_b}^2},
\end{equation}
where $\rho^*$ is the fluid density. Substituting Eq.~(\ref{eq:C_f}) into Eq.~(\ref{eq:E_p}), and denoting with $M^* = \rho^* V^*_f$ the total mass of the transported fluid, yields:
\begin{equation}
\label{eq:E_p2}
E^*_p = \frac{M^*{U^*_b}^2 C_f}{2A^*},
\end{equation}

\begin{figure}
\centering
\includegraphics[width=0.7\textwidth]{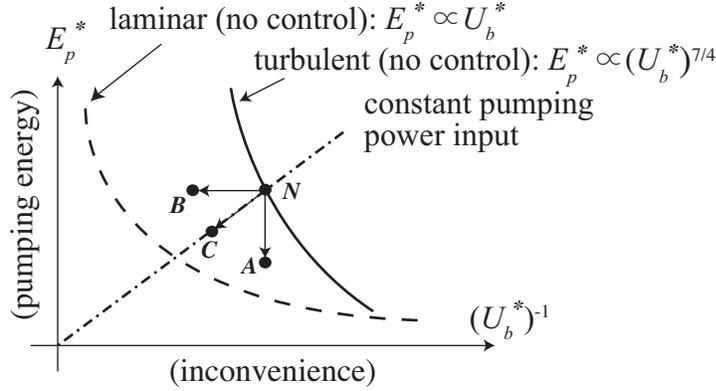}
\caption{Pumping energy $E^*_p$ versus the inverse of the bulk mean velocity $U^*_b$, which
reflects the time needed to pump a given amount of fluid through a duct with a given cross section and length.
Starting from the non-controlled flow state $N$, successful flow control under CFR shifts it to $A$, whereas successful flow control under CPG shifts it to $B$. Control at constant power input (CPI, see text) shifts $N$ to $C$.}
\label{fig:Ep_Ub}
\end{figure}

In order to evaluate control performance in terms of energy consumption and convenience, we start from the plot sketched in Fig.~\ref{fig:Ep_Ub}, where the vertical axis is pumping energy $E^*_p$ (and thus degree of energetic cost), and the horizontal axis is $1/{U^*_b}$, which represents the time for a fluid to travel over a unit length (and thus the degree of convenience). In a laminar flow $C_f \propto {U^*_b}^{-1}$ so that $E^*_p \propto U^*_b$, which is plotted as a dashed line in Fig.~\ref{fig:Ep_Ub}. In non-controlled turbulent flows, several empirical formulas exist that relate $C_f$ to $U_b^*$. For example, the Blasius correlation \citep{schlichting-1979} can be employed, from which $C_f \propto {U^*_b}^{-1/4}$ and $E^*_p \propto {U^*_b}^{7/4}$, which is depicted by a solid line in Fig.~\ref{fig:Ep_Ub}. The objective of turbulence control is to achieve a flow state located in the region left/below the solid line. This plot naturally emphasizes that reduced energy consumption can easily be achieved when one is willing to sacrifice convenience, i.e. wait longer for a certain amount of fluid to arrive, and that high convenience, i.e. extremely fast transport, increases the energy requirements significantly.

Suppose control is applied to the non-controlled flow state labeled as $N$ in Fig.~\ref{fig:Ep_Ub}.
When the bulk mean velocity is kept constant (CFR), the control shifts $N$ along the vertical arrow to, say, $A$.
The reduction of $E^*_p$, i.e. the distance $|NA|$ between points $N$ and $A$, is equivalent to drag reduction rate $R$. On the other hand, under the CPG condition, Eqns.~(\ref{eq:E_p}) and (\ref{eq:tau_dp}) indicate that $E^*_p$ also remains constant. Therefore, successful control shifts $N$ along the horizontal arrow to $B$. ($E_p^*$ is pumping energy per unit fluid volume: with drag reduction under CPG, pumping power is increased, but operation time is decreased, being inversely proportional to the flow rate. Energy, which is pumping power multiplied by operation time, is constant.)

The dash-dotted line connecting the origin and flow state $N$ represents the locus of points where the pumping {\em power}, i.e. the pumping energy divided by the operating time, remains constant. A shift to point $C$ along the arrow in Fig.~\ref{fig:Ep_Ub} therefore corresponds to a controlled state that requires the same power input as the non-controlled state (constant power input, CPI), while providing at the same time larger flow rate and smaller pumping energy.

\begin{figure}
\centering
\includegraphics[width=0.7\textwidth]{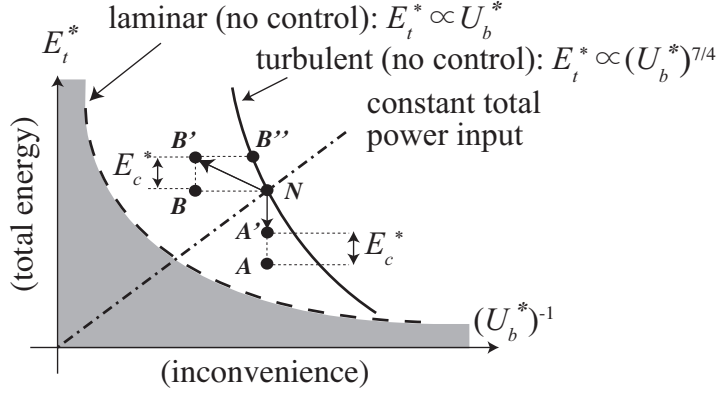}
\caption{Total energy $E^*_t$ versus the inverse of the bulk mean velocity $U^*_b$.
The vertical shifts from $A$ and $B$ in Fig.\ref{fig:Ep_Ub} to $A'$ and $B'$ correspond to
the energy $E^*_c$ required by the control technique.}
\label{fig:Et_Ub}
\end{figure}
If the flow control technique is of the active type and thus requires energy to operate, its energy input $E^*_c$ must enter the picture. In order to account for $E^*_c$, Fig.~\ref{fig:Ep_Ub} with just pumping energy $E^*_p$ is replaced by Fig.~\ref{fig:Et_Ub}, where the total energy $E^*_t = E^*_p + E^*_c$ is used on the vertical axis. The paths for a controlled flow state under constant flow rate and constant pressure gradient are shown by the arrows $NA'$ and $NB'$. The solid and broken lines for non-controlled turbulent and laminar flows are the same as those in Fig.~\ref{fig:Ep_Ub}, since $E^*_t = E^*_p$. The additional control energy input $E^*_c$ is reflected in Fig.~\ref{fig:Ep_Ub} by the shift of points $A$ and $B$ in the vertical direction to $A'$ and $B'$, respectively.
Of course, the two $E^*_c$ shown in Fig.~\ref{fig:Et_Ub} are not generally identical.
In this plot the distance $\left|NA\right|$, corresponds to the conventional drag reduction rate $R$ under CFR, while $\left|NA'\right|$ reflects the net energy savings $S$ and the ratio $\left|NA\right|/\left|A'A\right|$ the energy gain $G$. In a similar manner, $\left| NB \right|$ explains the decrease in $C_f$ achieved under CPG owing to the increase in flow rate, whereas $\left| B' B'' \right|$ quantifies the portion of that increase that is due to successful flow control and $\left| N B \right| - \left| B' B'' \right|$ is the remaining portion due to the increase in pumping energy.

According to \cite{bewley-2009} and \cite{fukagata-sugiyama-kasagi-2009}, the total energy consumption at a given flow rate is minimized when the flow becomes laminar. Therefore, no flow state can be located below the laminar curve, i.e., in the grey region in Fig.~\ref{fig:Et_Ub}. We also note that, in analogy to Fig.~\ref{fig:Ep_Ub}, a line connecting the origin in Fig.~\ref{fig:Et_Ub} and any point along the non-controlled curve represents flow states with the same total power consumption (CPI).

\begin{figure}
\centering
\includegraphics[width=0.9\textwidth]{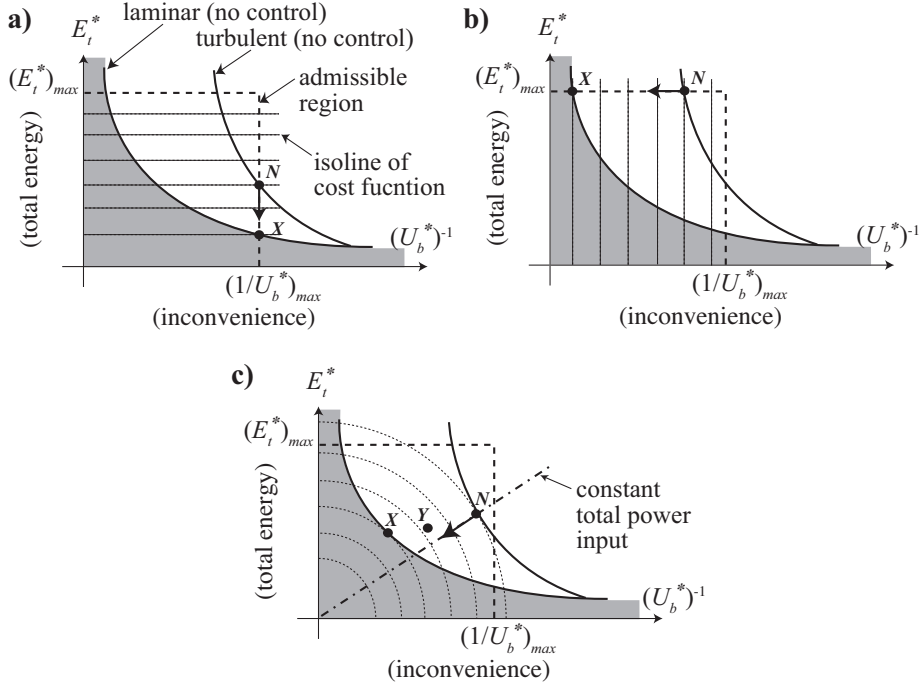}
\caption{An energy-convenience plane with isolines of typical cost functions $\mathcal{F}$ represented by dotted lines; a) $\mathcal{F}=E_t^*$, b) $\mathcal{F} = 1/U_b^*$, c) $\mathcal{F}= (E_t^*)^2 + (1/U_b^*)^2$.
An admissible region defined by $(E^*_t)_{max}$ and $(1/U_b^*)_{max}$  affordable in the application
is depicted by a square bounded by a dashed line. The flow state $N$ is optimal for uncontrolled flow, while the laminar state $X$ provides the minimum achievable value of the cost function by flow control.}
\label{fig:cost_function}
\end{figure}

By looking at the plane depicted in Fig.~\ref{fig:Et_Ub}, it becomes readily evident that several paths are possible to move the flow state from the non-controlled point $N$ towards the laminar curve. The three particular ones  considered so far, i.e. the CFR, CPG and CPI straight lines, are not the only ones, and not necessarily the best ones: only application-specific considerations allow the designer to favor one particular strategy. However, the energy-convenience plane is a natural workspace where the chosen strategy can be represented in terms of an application-dependent cost function $\mathcal{F}=\mathcal{F}(E_t^*, U_b^*)$ to be minimized.

Three energy-convenience maps with typical cost functions are shown in Fig.~\ref{fig:cost_function} a) - c).
The dotted lines represent the isolines of each cost function. The minimization is first constrained by the maximum affordable energy expenditure $(E^*_t)_{max}$ and by the maximum affordable inconvenience $(1/U_b^*)_{max}$.
This avoids a trivial solution such as $U_b^* = 0$ or $E^*_t=\infty$ when the cost function includes only one of the two competing factors, i.e., $E_t^*$ and $U_b^*$.
These constraints yield the admissible region bounded by a dashed line in Fig.~\ref{fig:cost_function}, and
the optimal flow state should exist either inside the region or on its boundary.

First, we consider the case where $\mathcal{F} = E_t^*$ as shown in Fig.~\ref{fig:cost_function} a).
When control is not applied, only the flow states on the uncontrolled turbulent curve are realizable.
In this case, point $N$ shown in Fig.~\ref{fig:cost_function} a) provides the minimum value of $\mathcal{F}$, and is therefore optimal.
The downward arrow from point $N$ shows the local gradient of $\mathcal{F}$, indicating that a control strategy changing the flow state in this direction
is most effective to decrease the objective function. This particular choice of $\mathcal{F}$ corresponds to the CFR condition. Since flow states below the laminar curve cannot be realized, the lower-bound of $\mathcal{F}$ is obtained at point $X$ on the laminar curve.

Similarly, when $\mathcal{F} = 1/U_b^*$, the optimal flow state without control is given by
the intersection point $N$ of the turbulent curve and the upper boundary of the admissible region
as shown in Fig.~\ref{fig:cost_function} b). In this case, the local gradient of $\mathcal{F}$ indicates that
enhancing $U_b^*$ under the constant $E_t^*$ is the optimal strategy.
If the applied control is passive or $E^*_c$ in active control is negligibly small,
$E^*_t$ is equivalent to $E^*_p$, so that the optimal control strategy results in the CPG condition.
Again, the minimum $\mathcal{F}$ is achieved at point $X$ on the boundary of the admissible region.

As a less obvious example, Fig.~\ref{fig:cost_function} c) shows isolines of $\mathcal{F} =  (E^*_t)^2 + (1/U_b^*)^2$, where the energy saving and the convenience are considered to be equally important.
The optimal flow state $N$ without control, as well as the optimal state $X$ with control, are located inside the admissible region; the local gradient of $\mathcal{F}$ is always pointing towards the origin,
indicating that the control under the CPI condition is locally the optimal strategy. However, this example also highlights that the minimum of $\mathcal{F}$ at $X$ is reached along a non-trivial curve that does not correspond to the global CPI constraint. In general, the possible maximum reduction of the cost function is given by $\mathcal{F}(N) - \mathcal{F}(X)$, and the performance  $\eta$ of a control that leads to a certain flow state $Y$ shown in Fig.~\ref{fig:cost_function} c) can be expressed with respect to its potential, by computing the ratio between the realized reduction of $\mathcal{F}$ and the maximum possible reduction, i.e.
\[
\eta = \frac{\mathcal{F}(N)-\mathcal{F}(Y)}{\mathcal{F}(N) - \mathcal{F}(X)} .
\]


\subsection{Non-dimensional analysis}

The $E^*_t-{U^*_b}^{-1}$ plane described above is useful to evaluate energy saving and convenience achieved in a given flow system by flow control; however, it is not universal, since the variables on both axes are dimensional. Its practical value is largely increased by introducing a non-dimensional version through appropriate hydrodynamic quantities.

The horizontal axis can be easily made dimensionless by using $\nu^*/(U^*_b D^*) = Re_D^{-1}$, where
$Re_D$ is the diameter-based Reynolds number and $\nu^*$ is the fluid kinematic viscosity. To deal with the vertical axis, we first introduce an {\em effective} wall friction ${\tau^e_w}^*$
based on the total power consumption $P^*_t = P^*_p + P^*_c$:
\begin{equation*}
{\tau_w^e}^* = \frac{P^*_t}{U^*_b} = \tau^*_w + \frac{P^*_c}{U^*_b}.
\end{equation*}

The effective friction ${\tau_w^e}^*$ reduces to the conventional friction $\tau_w^*$ in absence of active flow control; its higher values in actively controlled flows reflect the energetic cost of the control. Employing  Eq.~(\ref{eq:E_p2}), the total energy consumption $E^*_t$ is obtained by simply replacing $C_f$ with $C^e_f$ as:
\begin{equation}
\label{eq:E_t}
E^*_t = \frac{M^*{U^*_b}^2 C^e_f}{2A^*},
\end{equation}
where $C^e_f$ is the effective friction coefficient defined in analogy to (\ref{eq:C_f}) as:
\begin{equation}
\label{eq:C_f^e}
C^e_f  = \frac{{\tau^e_w}^*}{\frac{1}{2}\rho^* {U^*_b}^2}.
\end{equation}

Thus, the vertical axis may be interpreted as an effective friction coefficient:
\begin{equation}
C_f^e = \frac{2A^* E^*_t}{M^*{U^*_b}^2}.
\label{eq:cfe}
\end{equation}
and the plane described in Fig.~\ref{fig:Et_Ub} becomes analogous to the conventional $C_f-Re$ plane, with the added benefit of including the energetic cost of the control.

However, like in the usual $C_f - Re$ plane, the above form is still not suitable for the present purpose, since the measure of convenience, i.e., $U^*_b$, appears explicitly in the denominator of $C_f^e$. In order to avoid this, multiplication of Eq.(\ref{eq:cfe}) with $Re_D^2$ results in:
\begin{equation}
C_f^e Re^2_D = \frac{2A^* E^*_t}{M^* {(\nu^*/D^*)}^2}.
\end{equation}
This way, $E^*_t$ is non-dimensionalized by the fluid viscosity and geometrical properties of the duct only.

\begin{figure}
\centering
\includegraphics[width=\textwidth]{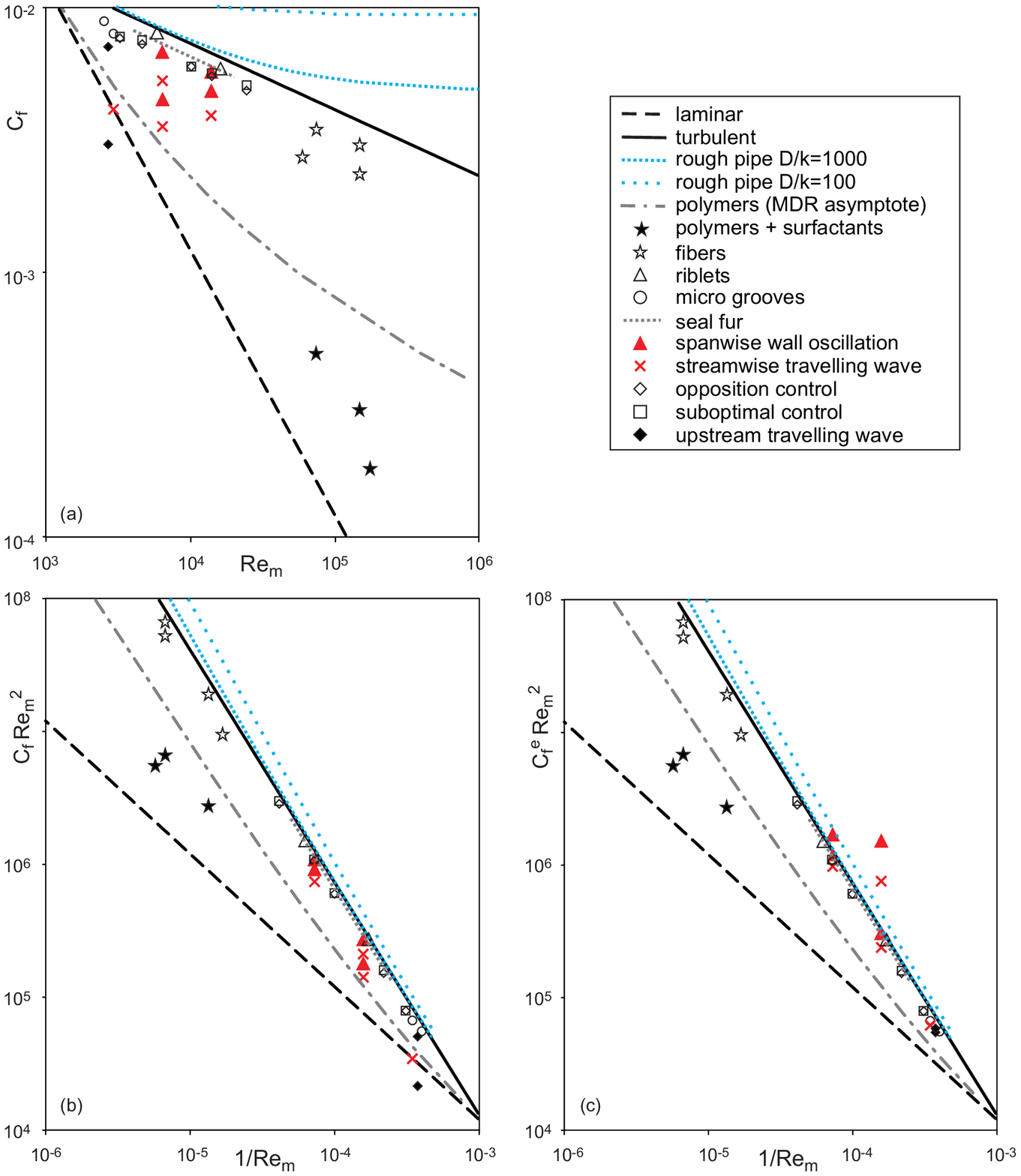}
\caption{Friction coefficient versus Reynolds number (a) including a number of available literature data \citep{colebrook-1939, virk-mickley-smith-1974, lee-vaseleski-metzner-1974, delfos-hoving-westerweel-2011, grueneberger-hage-2011, garcia-jimenez-2011, frohnapfel-jovanovic-delgado-2007, itoh-etal-2006b, quadrio-ricco-2004, quadrio-ricco-viotti-2009, iwamoto-suzuki-kasagi-2002, min-etal-2006} and its evolution towards  a nondimensional form of the suggested energy savings versus convenience plane (c) of Fig.~\ref{fig:Et_Ub} where (b) corresponds to the nondimensional form of Fig.~\ref{fig:Ep_Ub}. }
\label{fig:cf-Re}
\end{figure}

Fig.~\ref{fig:cf-Re}(a-c) graphically illustrates the transformation from the well established $C_f-Re$ plane to the non-dimensional version of the $E^*_t-{U^*_b}^{-1}$ plane, through the intermediate step of the $C_f Re^2-Re^{-1}$ plane. Plotted in Fig.~\ref{fig:cf-Re} are also a number of existing datasets, for which information about raw and net drag reduction are available. The Colebrook-White correlation \citep{colebrook-1939} for two different wall roughnesses is also shown. Since most of the drag reduction results correspond to channel flows, the bulk Reynolds number of a channel flow, $Re_m$, based on the full channel height and $U^*_b$, is used. For data obtained in pipe flows \citep{colebrook-1939, virk-mickley-smith-1974, lee-vaseleski-metzner-1974} the Reynolds number is corrected based on the ratio between the Dean correlation for a turbulent channel flow \citep{dean-1978} and Prandtl's universal law of friction for smooth pipes \citep{schlichting-1979}, to obtain a collapse of the corresponding correlations in the laminar and the turbulent regime.

Note first that the vertical axis of Fig.~\ref{fig:cf-Re}b) corresponds to the square of the von Karman number: $\mathrm{Ka}^2 = C_f Re_m^2 $. Historically, the Prandtl--von Karman plot, i.e. $1/\sqrt{f}$ vs. $Re\sqrt{f}=\mathrm{Ka}$ with $f=C_f$, has often been used to describe the relation between pressure drop and flow rate in a fluid system operated under CPG. The Virk asymptote for maximum drag reduction with polymer additives \citep{virk-mickley-smith-1974} is for example given in these coordinates. Except for proportionality constants, the Prandtl--von Karman plot corresponds to $U_b^+=U^*_b/u_\tau$ versus $Re_\tau=u_\tau H^*/ 2\nu^*$ for a duct with fixed geometric properties; this indicates that the energy input (given by the pressure gradient and thus the wall friction velocity $u_\tau=\sqrt{\tau^*_w/\rho^*}$) is used to obtain the values on both axes. In contrast, the dimensionless energy--convenience plane proposed here effectively decouples energy and flow rate on the two axes, and is therefore inherently superior in the context of flow control.

The final plane, i.e. that depicted in Fig.~\ref{fig:cf-Re}c), accounts for the total energy consumption through $C_f^e$. The curves representing laminar and turbulent flow states remain unchanged from Fig.~\ref{fig:cf-Re}b) to Fig.~\ref{fig:cf-Re}c), since $E_t^* = E_p^*$ and $C^e_f=C_f$; however, points corresponding to active flow control techniques are shifted upwards to account for the additional energy requirement.

In Fig.~\ref{fig:cf-Re} any point located below the solid line representing the non-controlled turbulent state corresponds to a case of successful flow control. The comparison between a) and b) with c) reveals that the term ``successful'' depends on whether drag reduction or energy saving is defined as the goal of the applied control. As an example, spanwise wall oscillations \citep{quadrio-ricco-2004} and streamwise traveling waves of spanwise motion \citep{quadrio-ricco-viotti-2009} can yield, for certain choices of parameters, large reductions of drag but negative net energy savings. These points are located below the turbulent line in part a) and b) and above the turbulent line in part c) of the figure. In Fig.~\ref{fig:cf-Re}a) and b) one point is located below the dashed line representing the laminar flow state, and corresponds to the result of \cite{min-etal-2006}, where sublaminar drag is realized by the introduction of an upstream traveling wave of blowing and suction. As in Fig.~\ref{fig:Et_Ub}, values below the laminar line are not allowed in Fig.~\ref{fig:cf-Re}c), and the point is shifted upwards accordingly when the energy used for the generation of the wave is taken into account.

The influence of wall-roughness can be evaluated in a similar manner: Assume that a certain amount of fluid has to be transported through a given duct with a given total power input. The intersection points of the objective function with the curves for different wall roughnesses indicate which additional cost (energy consumption) and time requirements arise due to a rougher surface in the duct, or which decrease in cost and improvement in time performance can be obtained by using a smoother surface.

\begin{figure}
\centering
\includegraphics[width=0.55\textwidth]{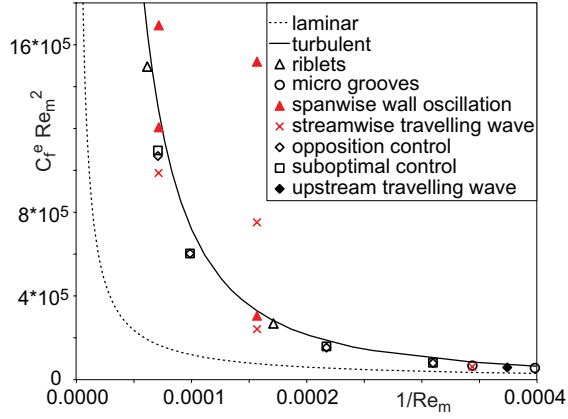}
\caption{Expanded energy-convenience map for the region of low Reynolds numbers (linear scale).}
\label{fig:cfRe2_expanded}
\end{figure}
Except for additives, data included in the plot are clustered at lower Reynolds number (where DNS and experiments are more easily carried out) and quite close to the turbulent line. In order to better reevaluate these results, a zoom of the energy-convenience map centered at the low Reynolds numbers is shown in Fig.~\ref{fig:cfRe2_expanded} with linear scale. With increasing convenience, i.e., the Reynolds number, the curve for the uncontrolled turbulent flow rapidly rises. This corresponds to an increase in the required pumping energy to drive the turbulent flow. Such a trend is difficult to extract from the conventional $C_f-Re$ plot, where $C_f$ monotonically (although gradually) decreases with increasing $Re$. It is also evident that the amount of energy saved by active control at higher Reynolds number is much larger than that at low Reynolds number. This is because the energy consumption shown in the ordinate is proportional to not only $C^e_f$ itself, but also $Re^2$. In particular, the control technique based on streamwise-traveling waves of spanwise wall velocity are observed to yield significant energy savings at relatively high values of $Re$.

\section{External flows}

\begin{figure}
\centering
\includegraphics[width=0.6\textwidth]{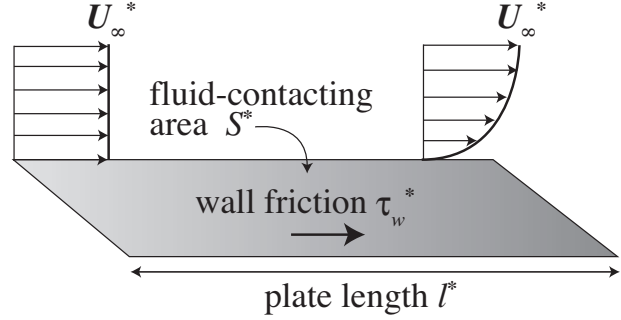}
\caption{Schematic of a flat plate moving at speed $U^*_{\infty}$ at zero angle of attack.}
\label{fig:flat_plate}
\end{figure}

The above discussion for internal flows can readily be extended to external flows, as long as we limit ourselves to considering friction drag.
In order to demonstrate the idea, a solid body moving through a quiescent fluid (an airplane, for example) is simplified to a flat plate which moves from point $A$ to point $B$ at zero angle of attack, where the distance between the two points is denoted as $L^*$. The traveling speed of the flat plate is $U^*_{\infty}$, while the streamwise length and total area of the flat plate are given by $\ell^*$ and $S^*$, respectively (see Fig.~\ref{fig:flat_plate}).
Since pressure drag does not arise on such an airfoil with infinitesimal thickness, the force $F^*$ acting on the moving plate is attributed to the friction drag only,
which is given by
\begin{equation}
F^* = \int_{S^*} \tau'^*_w ds,
\end{equation}
where $\tau'^*_w$ is the wall shear stress, which varies along the plate. By defining a local friction coefficient  as:
\begin{equation}
\label{eq:Cf_external}
C'_f = \frac{\tau'^*_w}{\frac{1}{2}\rho^* {U^*_{\infty}}^2}.
\end{equation}
the propulsion energy $E'^*_p$ per unit of fluid-contacting area required for the plate to travel a unit distance can be written as:
\begin{equation}
E'^*_p = \frac{1}{S^*}\int_{S^*} \tau'^*_w ds = \overline{\tau'^*_w} = \frac{1}{2}\rho^*{U^*_{\infty}}^2 \overline{C'_f}.
\end{equation}
where the over-bar denotes the spatial average over the plate. 

Following the discussion for internal flows, convenience is defined as the traveling time per unit distance, i.e. $1/U^*_{\infty}$, so that the corresponding dimensionless parameter is $1/Re_\ell$, where the Reynolds number of a flat plate is defined as $Re_\ell = U^*_\infty \ell^*/\nu^*$. Similarly, the dimensionless parameter for energy consumption is defined as:
\begin{equation}
\overline{C'_f} Re^2_\ell = \frac{E'^*_p}{\frac{1}{2}\rho^*{U^*_{\infty}}^2} 
                                     \left( \frac{U^*_{\infty}\ell^*}{\nu^*} \right)^2 
                                     =  \frac{E'^*_p {\ell^*}^2}{\frac{1}{2}\rho^*{\nu^*}^2} 
\end{equation}
Therefore, the $C_f Re^2$-$ Re^{-1}$ plane introduced for internal flows
in the previous subsection can also be used for external flows.
In order to take into account the energy consumption for control, we again introduce the equivalent wall friction ${\tau'^e_w}^*$ as:
\begin{equation}
\overline{{\tau'^e_w}^*} = E'^*_t = E'^*_p + E^*_c =\overline{ \tau'^*_w} + E^*_c.
\end{equation}

Hence, by replacing $\tau'^*_w$ with ${\tau'^e_w}^*$ in Eq.~(\ref{eq:Cf_external}),
we can easily extend the $C^e_f Re^2 - Re^{-1}$ plane to external flows.

\section{Concluding discussion}
\label{sec:summary}
The optimization problem of flow control involves an interplay between energy saving and convenience, or -- more generally -- money and time. Starting from this observation, a methodology for assessing flow control techniques for skin-friction drag reduction is proposed. We derive two dimensionless parameters, i.e., $C_f^e Re^2$ and $Re^{-1}$, which express the cost of the total energy consumption and the convenience for transporting a fluid through a duct with a certain cross-sectional geometry. Any controlled flow state can be represented in the two-dimensional plane composed by
these two dimensionless quantities, without the need of imposing a constraint on the flow condition. The theoretical lower bound of the total energy consumption under a constant flow rate derived by \cite{bewley-2009} and \cite{fukagata-sugiyama-kasagi-2009} is naturally integrated into the plot.

The suggested ``energy-convenience plane'', which can also be used for external flows, extends the comparison of flow control techniques beyond the constant flow rate approach often used in literature up to now \citep{hoepffner-fukagata-2009, kasagi-hasegawa-fukagata-2009}, and allows the inclusion of application-specific cost functions such that the control performance can be judged in respect to a specific application. Practically, the overall cost should account for not only the energy consumption for pumping and control, but also fabrication, implementation and maintenance of sensors, actuators, polymers, surface structures and so forth. Since the vertical axis in the energy-convenience plane is defined as the energy consumption per unit fluid volume, additional initial and maintenance costs per unit fluid volume can be considered in a straightforward manner.The presented framework allows designers to easily include the energetic costs due to fluid transport into a budget and thus decide which target flow state provides the best economical choice for a specific application.

\section*{Acknowledgements}
The authors acknowledge support of projects FR2823/2-1 and 3-1 of the German Research Foundation (DFG) and the Cluster of Excellence``Center of Smart Interfaces'' at TU Darmstadt. Yosuke Hasegawa greatly acknowledges the support from the Japan Society for the Promotion of Science (JSPS) Postdoctoral Fellowship for Research Abroad.

\bibliographystyle{jfm}

\begin{thebibliography}{24}
\expandafter\ifx\csname natexlab\endcsname\relax\def\natexlab#1{#1}\fi

\bibitem[Bewley(2009)]{bewley-2009}
{\sc Bewley, T.R.} 2009 A fundamental limit on the balance of power in a
  transpiration-controlled channel flow. {\em J. Fluid Mech.\/} {\bf 632},
  443--446.

\bibitem[Choi {\em et~al.\/}(1994)Choi, Moin \& Kim]{choi-moin-kim-1994}
{\sc Choi, H., Moin, P. \& Kim, J.} 1994 Active turbulence control for drag
  reduction in wall-bounded flows. {\em J. Fluid Mech.\/} {\bf 262}, 75--110.

\bibitem[Colebrook(1939)]{colebrook-1939}
{\sc Colebrook, C.F.} 1939 Turbulent flows in pipes with particular reference
  to the transition between the smooth- and rough-pipe laws. {\em J. Inst.
  Civil Eng.\/} {\bf 11}, 133--156.

\bibitem[Dean(1978)]{dean-1978}
{\sc Dean, R.B.} 1978 Reynolds number dependence of skin friction and other
  bulk flow variables in two-dimensional rectangular duct flow. {\em Trans.
  ASME I: J. Fluids Eng.\/} {\bf 100}, 215.

\bibitem[Delfos {\em et~al.\/}(2011)Delfos, Hoving, Westerweel \&
  Boersma]{delfos-hoving-westerweel-2011}
{\sc Delfos, R., Hoving, J., Westerweel, J. \& Boersma, B.J.} 2011 Experiments
  on drag reduction by fibers in turbulent flows. In {\em Euromech meeting 513:
  {D}ynamics of non-spherical particles in fluid turbulence\/}. April 6-8 2011,
  Udine (I).

\bibitem[Frohnapfel {\em et~al.\/}(2007)Frohnapfel, Jovanovi\'c \&
  Delgado]{frohnapfel-jovanovic-delgado-2007}
{\sc Frohnapfel, B., Jovanovi\'c, J. \& Delgado, A.} 2007 Experimental
  investigation of turbulent drag reduction by surface-embedded grooves. {\em
  J. Fluid Mech.\/} {\bf 590}, 107--116.

\bibitem[Fukagata {\em et~al.\/}(2002)Fukagata, Iwamoto \&
  Kasagi]{fukagata-iwamoto-kasagi-2002}
{\sc Fukagata, K., Iwamoto, K. \& Kasagi, N.} 2002 Contribution of {R}eynolds
  stress distribution to the skin friction in wall-bounded flows. {\em Phys.
  Fluids\/} {\bf 14}~(11), L73--L76.

\bibitem[Fukagata {\em et~al.\/}(2009)Fukagata, Sugiyama \&
  Kasagi]{fukagata-sugiyama-kasagi-2009}
{\sc Fukagata, K., Sugiyama, K. \& Kasagi, N.} 2009 On the lower bound of net
  driving power in controlled duct flows. {\em Physica D\/} {\bf 238},
  1082--1086.

\bibitem[Garcia-Mayoral \& Jim\'enez(2011)]{garcia-jimenez-2011}
{\sc Garcia-Mayoral, R. \& Jim\'enez, J.} 2011 Hydrodynamic stability and the
  breakdown of the viscous regime over riblets. {\em J. Fluid. Mech.\/} {\bf
  678}, 317--347.

\bibitem[Gr\"uneberger \& Hage(2011)]{grueneberger-hage-2011}
{\sc Gr\"uneberger, R. \& Hage, W.} 2011 Drag characteristics of longitudinal
  and transverse riblets at low dimensionless spacings. {\em Exp. Fluids\/}
  {\bf 50}~(2), 363--373.

\bibitem[H{\oe}pffner \& Fukagata(2009)]{hoepffner-fukagata-2009}
{\sc H{\oe}pffner, J. \& Fukagata, K.} 2009 Pumping or drag reduction? {\em J.
  Fluid Mech.\/} {\bf 635}, 171--187.

\bibitem[Itoh {\em et~al.\/}(2006)Itoh, Tamano, Iguchi, Yokota, Akino, Hino \&
  Kubo]{itoh-etal-2006b}
{\sc Itoh, M., Tamano, S., Iguchi, R., Yokota, K., Akino, N., Hino, R. \& Kubo,
  S.} 2006 Turbulent drag reduction by the seal fur surface. {\em Phys.
  Fluids\/} {\bf 18}~(065102), 1--9.

\bibitem[Iwamoto {\em et~al.\/}(2002)Iwamoto, Suzuki \&
  Kasagi]{iwamoto-suzuki-kasagi-2002}
{\sc Iwamoto, K., Suzuki, Y. \& Kasagi, N.} 2002 Reynolds number effect on wall
  turbulence: toward effective feedback control. {\em Int. J. Heat Fluid
  Flow\/} {\bf 23}, 678--689.

\bibitem[Jung {\em et~al.\/}(1992)Jung, Mangiavacchi \&
  Akhavan]{jung-mangiavacchi-akhavan-1992}
{\sc Jung, W.J., Mangiavacchi, N. \& Akhavan, R.} 1992 Suppression of
  turbulence in wall-bounded flows by high-frequency spanwise oscillations.
  {\em Phys. Fluids A\/} {\bf 4 (8)}, 1605--1607.

\bibitem[Kasagi {\em et~al.\/}(2009)Kasagi, Hasegawa \&
  Fukagata]{kasagi-hasegawa-fukagata-2009}
{\sc Kasagi, N., Hasegawa, Y. \& Fukagata, K.} 2009 Towards cost-effective
  control of wall turbulence for skin-friction drag reduction. {\em Advances in
  Turbulence XII\/}, vol. 132, pp. 189--200. Springer.

\bibitem[Lee {\em et~al.\/}(1974)Lee, Vaselesky \&
  Metzner]{lee-vaseleski-metzner-1974}
{\sc Lee, W.K., Vaselesky, R.C. \& Metzner, A.B.} 1974 Turbulent drag reduction
  in polymerics solutions containing suspended fibers. {\em AIChE Journal\/}
  {\bf 20}~(1), 128--133.

\bibitem[Marusic {\em et~al.\/}(2007)Marusic, Joseph \&
  Mahesh]{marusic-joseph-mahesh-2007}
{\sc Marusic, I., Joseph, D.~D. \& Mahesh, K.} 2007 Laminar and turbulent
  comparisons for channel flow and flow control. {\em J. Fluid Mech.\/} {\bf
  570}, 467--477.

\bibitem[Min {\em et~al.\/}(2006)Min, Kang, Speyer \& Kim]{min-etal-2006}
{\sc Min, T., Kang, S.M., Speyer, J.L. \& Kim, J.} 2006 Sustained sub-laminar
  drag in a fully developed channel flow. {\em J. Fluid Mech.\/} {\bf 558},
  309--318.

\bibitem[Quadrio \& Ricco(2004)]{quadrio-ricco-2004}
{\sc Quadrio, M. \& Ricco, P.} 2004 Critical assessment of turbulent drag
  reduction through spanwise wall oscillation. {\em J. Fluid Mech.\/} {\bf
  521}, 251--271.

\bibitem[Quadrio \& Ricco(2011)]{quadrio-ricco-2011}
{\sc Quadrio, M. \& Ricco, P.} 2011 The laminar generalized {S}tokes layer and
  turbulent drag reduction. {\em J. Fluid Mech.\/} {\bf 667}, 135--157.

\bibitem[Quadrio {\em et~al.\/}(2009)Quadrio, Ricco \&
  Viotti]{quadrio-ricco-viotti-2009}
{\sc Quadrio, M., Ricco, P. \& Viotti, C.} 2009 Streamwise-traveling waves of
  spanwise wall velocity for turbulent drag reduction. {\em J. Fluid Mech.\/}
  {\bf 627}, 161--178.

\bibitem[Schlichting(1979)]{schlichting-1979}
{\sc Schlichting, H.} 1979 {\em Boundary-layer theory\/}. McGraw Hill.Inc.

\bibitem[Spalart \& McLean(2011)]{spalart-mclean-2011}
{\sc Spalart, P.R. \& McLean, J.D.} 2011 Drag reduction: enticing turbulence,
  and then an industry. {\em Phil. Trans. R. Soc. A\/} {\bf 369}~(1940),
  1556--1569.

\bibitem[Virk {\em et~al.\/}(1974)Virk, Mickley \&
  Smith]{virk-mickley-smith-1974}
{\sc Virk, P.S., Mickley, H.S. \& Smith, K.A.} 1974 The ultimate asymptote and
  mean flow structure in {T}om's phenomenon. {\em J. Appl. Mech.\/} {\bf
  37}~(2), 488--493.

\end{thebibliography}

\end{document}